\documentclass[aps,pra,twocolumn,superscriptaddress]{revtex4}

\usepackage{epsfig}
\usepackage{amsmath,amssymb,amsfonts}

\newcommand{\beq}{\begin{equation}}
\newcommand{\eeq}{\end{equation}}
\newcommand{\bal}{\begin{align}}
\newcommand{\eal}{\end{align}}

\newcommand{\ket}[1]{\left|{#1}\right\rangle}
\newcommand{\bra}[1]{\left\langle{#1}\right|}
\newcommand{\braket}[2]{\left\langle{#1}\Big|{#2}\right\rangle}
\newcommand{\tr}[1]{{\rm tr}\left\{#1\right\}}

\begin{document}

\title{Symmetric minimal quantum tomography by successive measurements}
\author{Amir Kalev}
\affiliation{Centre for Quantum Technologies, National University of Singapore, 3 Science Drive 2, 117543, Singapore}
\author{Jiangwei Shang}
\affiliation{Centre for Quantum Technologies, National University of Singapore, 3 Science Drive 2, 117543, Singapore}
\author{Berthold-Georg Englert}
\affiliation{Centre for Quantum Technologies, National University of Singapore, 3 Science Drive 2, 117543, Singapore}
\affiliation{Department of Physics, National University of Singapore, 2 Science Drive 3, 117542, Singapore}
\date{March 08, 12}
\begin{abstract}
We consider the implementation of a symmetric informationally complete probability-operator measurement (SIC POM) in the Hilbert space of a $d$-level system by a two-step measurement process: a diagonal-operator measurement with high-rank outcomes, followed by a rank-1 measurement in a basis chosen in accordance with the result of the first measurement. We find that any Heisenberg-Weyl group-covariant SIC POM can be realized by such a sequence where the second measurement is simply a measurement in the Fourier basis, independent of the result of the first measurement. Furthermore, at least for the particular  cases studied, of dimension 2, 3, 4, and 8, this scheme reveals an unexpected operational relation between mutually unbiased bases and SIC POMs; the former are used to construct the latter. As a laboratory application of the two-step measurement process, we propose  feasible optical experiments that would realize SIC POMs in various dimensions. 
\end{abstract}
\maketitle

\section{Introduction}
Quantum state tomography is a measurement procedure designed to acquire complete information about the state of a given quantum system. It is an important component in most, if not all, quantum computation and quantum communication tasks.  The successful execution of such tasks hinges in part on the ability to  assess the state of the system at various stages. 

A general measurement in quantum mechanics is a probability-operator measurement (POM). A POM is informationally complete (IC) if any state of the system is determined completely by the measurement statistics \cite{ic1,ic2,ic3}. State tomography infers these probabilities from the data acquired with the aid of the POM. 

A symmetric IC POM (SIC POM) is an IC POM of a particular kind. In a $d$-dimensional Hilbert space (of kets) it is composed of $d^2$ subnormalized rank-1 projectors, $\{{\cal P}_j\}_{j=1}^{d^2}$, with equal pairwise fidelity of $1/(d+1)$. Their high symmetry and high tomographic efficiency have attracted the attention of many researchers, and a lot of work, both analytical and numerical, has been devoted to the construction of SIC POMs  in various dimensions, see e.g. \cite{berge04,renes04,appleby05,scott06,scott10,zauner11}. 

A group-covariant SIC POM is a measurement which  can be generated from a single projector (onto the so-called fiducial state) under the action of a group consisting of unitary operations.  Almost all known SIC POMs are covariant with respect to the Heisenberg-Weyl (HW) group (also known as the generalized  Pauli group). Two SIC POMs are said to be equivalent if there is a unitary operator that maps one SIC POM to the other.

In contrast to the  major theoretical progress all experiments and even proposals for experiments implementing  SIC POMs have been limited to the very basic quantum system, the two-level system (qubit) \cite{ling06,pimenta10}, with the exception of the recent experiment by Medendorp {\it et al.} \cite{steinberg11} where a SIC POM for a three-level system was approximated. This is, in part, due to the fact that there is no systematic procedure for implementing SIC POMs in higher dimensions, in a simple experimental set-up. 

Recently, however, an experiment that realizes a SIC POM in the four-dimensional Hilbert space of a qubit pair was proposed \cite{letter}. The experimental scheme exploited a new approach to SIC POMs that uses a two-step process: a measurement with full-rank outcomes, followed by a projective measurement on a basis that is chosen in accordance with the result of the first measurement. 
%
%
In this work, following the ideas presented in Ref. \cite{letter}, we explore the possibilities to implement SIC POMs using a successive-measurement scheme. We start by `breaking' a given SIC POM into two successive measurements, each with $d$ outcomes, with the intention that each measurement would be relatively easy to implement. Unexpectedly, we find that this approach provides a simple, systematic procedure to implement all HW group-covariant SIC POMs (HW SIC POMs). The latter could be realized by first implementing a POM with high-rank outcomes diagonal in a given basis  followed by a rank-1 projective measurement, where the basis of the first measurement and the basis of the second measurement are related by the Fourier transform. 

Based on this approach, we propose a general experimental scheme implementing HW SIC POMs in the Hilbert space of a $d$-dimensional quantum system (a qudit). In this scheme, the qudit is carried by a single photon as a path qudit, and the implementation is accomplished with the means of linear optics. In particular, we show that the one-parameter family of non-equivalent HW SIC POMs in dimension 3, could be implemented using the successive measurement approach in a single experimental set-up. Furthermore, we study the construction of the known SIC POMs in dimensions 2, 4, and 8 from two successive measurements. We find that the concept of mutually unbiased bases (MUB) plays a central role in the construction of SIC POMs in these dimensions --- a hint at a possibly profound link between SIC POMs and MUB;  Ref.~\cite{durt10} is a recent review on MUB.

The paper is organized as follows. Section~\ref{general} is concerned with finite-dimensional Hilbert spaces. There we discuss the formulation of SIC POMs in general, and the HW SIC POMs in particular, in terms of two successive measurements. Then we study the construction of known SIC POMs in particular dimensions. In Sec.~\ref{dim2}, we reformulate the SIC POM  in dimension 2 (known as the tetrahedron measurement) in terms of successive measurements, and show that the actual implementation of it by Ling {\it et al.} \cite{ling06} was indeed carried out using a sequential-measurement scheme. We also show how a relation between the SIC POM and MUB in dimension 2 is revealed through this formulation. In Sec.~\ref{dim3} we study the decomposition of all known non-equivalent SIC POMs in dimension 3 into two successive measurements. We show that this decomposition allows the implementation of all (known) non-equivalent SIC POMs in dimension 3 with a single experimental set-up. In Sec.~\ref{dim4}, we study the realization of the (known) SIC POMs in dimension 4 by successive measurements. Here we also find an interesting, structural and operational relation between MUB and SIC POMs. We briefly  describe a proposal for their implementation, using single photon sources together with passive linear optical elements \cite{letter}.  In Sec.~\ref{dim8}, we discuss the construction of the three known, non-equivalent, group-covariant SIC POMs in dimension 8 in terms of successive measurements. We show that the one that is covariant with respect to the 3-qubit Pauli group has the same structure as the SIC POMs in the other studied dimensions. Finally, we offer conclusions in Sec.~\ref{conclusions}.
\section{The general case}\label{general}
A general measurement on a quantum system is composed of a set of outcomes. The latter are mathematically represented by positive operators ${\cal P}_j$ that sum up to the identity operator.  The probability of obtaining the outcome ${\cal P}_j$ is given by the Born rule:  \mbox{$p_j = \tr{{\cal P}_j\rho}$}, where $\rho$ is the pre-measurement statistical operator of the system. If the $j$th outcome is found, the post-measurement statistical operator of the system is given by
\mbox{$\rho_{j}=\frac{1}{p_j}P_j\rho P^\dagger_j$}, 
where $P_j$ is the relevant Kraus operator for the $j$th outcome, \mbox{${\cal P}_j=P^\dagger_j P _j$}. Note that the decomposition of the ${\cal P}$s into the corresponding Kraus operators is not unique; for example, \mbox{$P^\dagger_j P_j$} is invariant under the unitary transformation \mbox{$P_j\rightarrow U_jP_j$}, with different $U_j$s corresponding to different implementations of the POM.

Suppose that a given system is subjected to a sequence of two POMs, each with $d$ outcomes,  \mbox{$\{{\cal A}_k=A^\dagger_k A_k\}_{k=1}^{d}$}, followed by \mbox{$\{{\cal B}^{\scriptscriptstyle(k)}_j\}_{j=1}^{d}$}, where the superscript $k$ indicates that in general the second measurement depends on the actual outcome of the first measurement. Following Born's rule, the probability of obtaining the $n$th and $m$th outcomes for the first and second measurements is given by \mbox{$\tr{\rho A^\dagger_n {\cal B}^{\scriptscriptstyle(n)}_mA_n}$}. Accordingly, the two successive measurements are equivalent to a single POM with $d^2$ outcomes \mbox{${\cal P}_{n,m}=A^\dagger_n {\cal B}^{\scriptscriptstyle(n)}_mA_n$} with \mbox{$n,m=1,\;\ldots,\;d$}. Indeed, summing \mbox{${\cal P}_{n,m}$} over the outcomes labeled by $m$ yields the outcome ${\cal A}_n$. Therefore,  upon finding the over-all outcome ${\cal P}_{n,m}$, we know that the $n$th outcome of the first POM and the $m$th outcome of the second POM are the case. In what follows we will identify the $A$s and the ${\cal B}$s such that the ${\cal P}$s make up a SIC POM in the $d$-dimensional Hilbert space of a qudit.

\subsection{HW SIC POMs}
Let us begin by showing that {\it all} HW SIC POMs could be realized by a two-step measurement scheme with a rather simple structure --- a high-rank, diagonal-operator measurement, followed by a measurement in the Fourier basis. In a $d$-dimensional Hilbert space, the HW group is composed of $d^2$ unitary operators \mbox{$\{X^kZ^j\}$}, \mbox{$j,k=1,\ldots,d$}, with $Z$ and $X$, the generators of the group, defined as
\beq
Z=\sum_{n=0}^{d-1}{\ket{n}\omega^n\bra{n}},\; X=\sum_{n=0}^{d-1}{\ket{n\oplus 1}\bra{n}},
\eeq
where \mbox{$\omega=e^{2\pi i/d}$} is the fundamental $d$th root of unity and $\oplus$ stands for the sum modulo $d$.
A HW SIC POM in a $d$-dimensional Hilbert space has $d^2$ outcomes, $\mathcal{P}_{k,j}$  \mbox{$k,j=1,\ldots,d$}, which are generated from a single projector onto a fiducial state, $\ket{\psi_{\rm fid}}$, under the action of the group elements,
\beq\label{hw}
\mathcal{P}_{k,j}=X^kZ^j\ket{\psi_{\rm fid}}\frac1{d}\bra{\psi_{\rm fid}}Z^{j\dagger}X^{k\dagger}.
\eeq
The fiducial state is chosen such that the $\mathcal{P}$s satisfy the defining property of a SIC POM,
\beq\label{tr}
\tr{\mathcal{P}_{k,j}\mathcal{P}_{m,n}}=\frac1{d^2}\left( \delta_{k,m}\delta_{j,n}+\left(1-\delta_{k,m}\delta_{j,n}\right)\frac1{d+1}\right).
\eeq
With the (normalized) fiducial state
\beq\label{psifid}
\ket{\psi_{\rm fid}}=\sum_{n=0}^{d-1}\ket{n}\alpha_n
\eeq
in Eq.~\eqref{hw}, we obtain
\beq\label{hw weak}
\mathcal{P}_{k,j}=\frac1{d}\sum_{n,m=0}^{d-1}\ket{m\oplus k}\alpha_m\omega^{(m-n)j}\alpha^*_n\bra{n\oplus k}.
\eeq
At this point we note that the right-hand-side of Eq.~\eqref{hw weak} has a two-step measurement structure. The Kraus operators corresponding to the outcomes of the first measurement are
\beq\label{a}
A_k=\sum_{m=0}^{d-1}\ket{m\oplus k}\alpha_m\bra{m\oplus k}
\eeq
with  \mbox{$k=1,\ldots,d$}, while the outcomes of the second measurement are projections onto the eigenstates of the Fourier  basis,
\beq\label{b}
{\cal B}_j=\frac1{d}\sum_{m,n=0}^{d-1}\ket{m}\omega^{(m-n)j}\bra{n},
\eeq
with  \mbox{$j=1,\ldots,d$}. Indeed, for Eqs.~\eqref{hw weak}-\eqref{b} we have
\beq\label{succHW}
\mathcal{P}_{k,j}=A^\dagger_k {\cal B}_jA_k,
\eeq
so that the HW SIC POM  for the fiducial state of Eq.~\eqref{psifid} is realized as a two-step measurement. This demonstrates the case.  If we relax the requirement that the sequential measurements in Eq.~\eqref{succHW} compose a {\it symmetric} IC POM, one can show  \cite{carmeli11} that in any finite-dimensional Hilbert space there exist $\alpha$s such that these measurements are IC.
\begin{figure}[t]
\centering
\includegraphics[scale=1.]{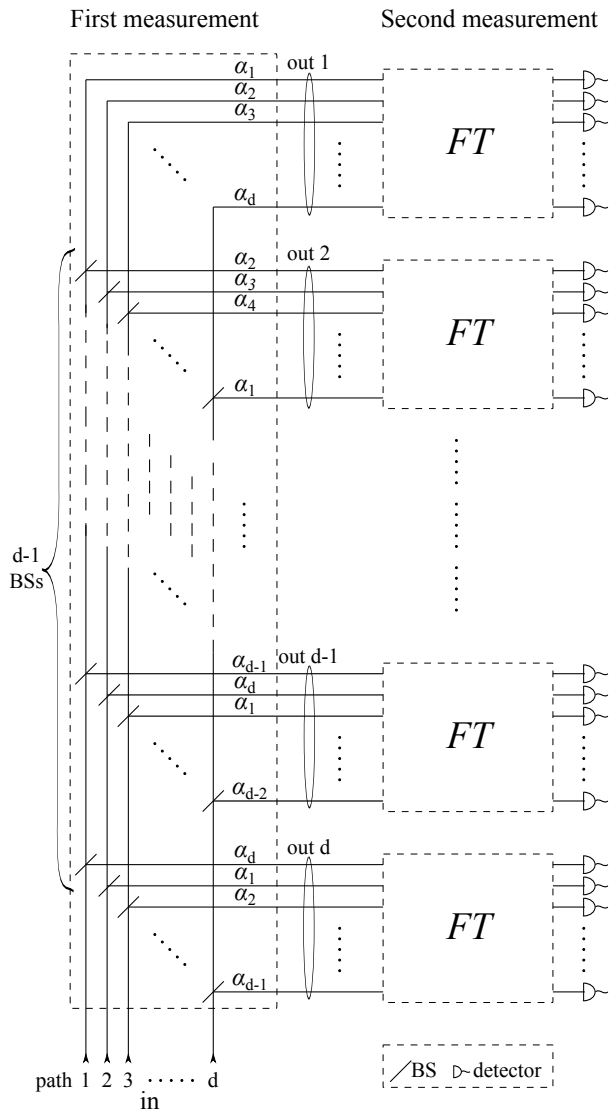}
\caption{An optical implementation of a HW SIC POM using a two-step measurement process.}
\label{fig:sicpom_hw}
\end{figure}

The mathematical formulation of SIC POMs as a two-step measurement process hints at the possibility for their implementation. Here, we propose a general experimental scheme with which any HW SIC POM in a $d$-dimensional Hilbert space of a qudit could be realized.  The qudit is carried by a single photon and is encoded in $d$ spatial alternatives of the photon (``path qudit''). A unitary transformation on the qudit state amounts to sending the photon through a set of beam splitters (BSs) and phase shifters (PSs), similar to the methods presented in~\cite{englert01}.

In this optical setting the HW SIC POMs are implemented as follows (see Fig.~\ref{fig:sicpom_hw} as a reference): The first measurement set-up is designed to implement the Kraus operators of Eq.~\eqref{a} by appropriately choosing the reflection amplitudes of the \mbox{$d-1$} BSs at each path.  The choice \mbox{$t_{n-1,k}r_{n,k}=\alpha_{k\ominus n}$}, where  $r_{n,k}$ and $t_{n,k}$ are the  reflection and the transmission amplitudes of the $n$th BS at the $k$th path (here \mbox{$n=1,\ldots,d-1$}, \mbox{$k=1,\ldots,d$}, the BS are counted from the entrance port, and the paths are numbered from left to right, as indicated in the figure; we define \mbox{$t_{0,k}=1$} for all $k$), ensures that a photon which enters the apparatus with a path statistical operator $\rho$, exits at port $k$  with the statistical operator \mbox{$A_k\rho A_k^\dagger/\tr{A_k\rho A_k^\dagger}$}. Upon exiting the first measurement apparatus, the photon is measured in the Fourier basis (indicated in the figure as a black-box labeled by $FT$). This measurement could be realized by a collection of BSs and appropriate PSs~\cite{reck94}.

\subsection{Fuzzy measurements}
So far, we considered the decomposition of a given HW SIC POM for a $d$-level system into a succession of two POMs, each with $d$ outcomes. Now, we would like to follow the reverse path, namely to start with a given structure for the two POMs, and study under what conditions they compose a SIC POM when measured in succession. In particular, we consider the situation where the first measurement is a `fuzzy measurement', where `fuzzy' means that each of the measurement outcomes corresponds to a projector onto the computational basis \mbox{$Z_k=\ket{k}\!\bra{k}$}, \mbox{$k=1,\ldots, d$}, mixed with the identity operator,
\beq\label{fuzzy}
{\cal A}_k=\frac1{d}(1-\lambda)+\lambda Z_k,
\eeq
whose positivity requires that \mbox{$\frac{-1}{d-1}\leq\lambda\leq 1$}.
Up to a unitary transformation, the corresponding Kraus operators  are 
\beq\label{krauss}
A_k=\sqrt{\frac{1-\lambda}{d}}\sum_{j(\neq k)}Z_j+\sqrt{\frac{1+(d-1)\lambda}{d}} Z_k.
\eeq
For the second measurement, we consider a projective measurement on a basis that is chosen in accordance with the result of the first measurement,
\beq\label{2nd}
{\cal B}^{(k)}_j= U^\dagger_k Z_j U_k\equiv{\textstyle\ket{j^{(k)}}\bra{j^{(k)}}},
\eeq
where $\ket{j^{(k)}}$ is the $j$th state of the $k$th basis. The unitary operator $U_k$ specifies the basis for the second measurement. It is worth recalling that  the outcomes of the first measurement are invariant under unitary transformations, \mbox{$A_k\rightarrow U_kA_k$}.  Therefore we can write the over-all outcomes of the two-step measurement process as
\beq\label{general pom}
{\cal M}_{k,j}= A^\dagger_k {\cal B}^{(k)}_j A_k.
\eeq

We now write the necessary and sufficient conditions that the ${\cal M}$s represent a SIC POM, that is: that this ansatz works. Equations~\eqref{tr},~\eqref{krauss}-\eqref{general pom} jointly require that 
\beq\label{cond1}
\lambda=\pm \frac{1}{\sqrt{1+d}},\;{\rm and}\;\;\left|\braket{m}{n^{(m)}}\right|^2=\frac1{d},
\eeq
as well as for \mbox{$k\neq m$}, 
\beq\label{cond2}
\left|\!\bra{n^{(m)}}\!\!\left[\alpha+(\beta-\alpha)\Big(\!\ket{m}\!\bra{m}\!+\!\ket{k}\!\bra{k}\!\Big)\!\right]\!\!\ket{j^{(k)}}\!\right|^2\!\!=\frac1{d},
\eeq
with \mbox{$\alpha=\sqrt{1-\lambda}$}, \mbox{$\beta=\sqrt{1+\lambda(d-1)}$}, and all indices take on the values \mbox{$1,\ldots ,d$}.  From the condition on $\lambda$ given right after Eq.~\eqref{fuzzy}, we get that \mbox{$\lambda=1/\sqrt{1+d}$} for \mbox{$d\geq 4$}. Note that the indices $k$ and $m$ label the first measurement while $j$ and $n$ label the second one. Recalling the definition of unbiased bases \cite{durt10}, Eq.~\eqref{cond1} implies that different bases of the second measurement are unbiased to one of the states from the computational basis.

We are able to solve Eqs.~\eqref{cond1} and \eqref{cond2} for dimension 2 and 3, and can also show that the known SIC POM in dimension 4 is a solution for these equations. Unfortunately we did not manage to solve, or prove the existence of a solution, for these equations in higher dimensions. In the following sections we discuss the solutions for these equations.

\section{Dimension 2: A qubit}\label{dim2}
The solution for Eqs.~\eqref{cond1} and \eqref{cond2} in dimension 2 is fairly straight-forward since  \mbox{$\ket{m}\bra{m}+\ket{k}\bra{k}=1$} for \mbox{$k\neq m$}. Accordingly, for \mbox{$\displaystyle{\lambda =\pm 1/\sqrt{3}}$}, Eq.~\eqref{cond2} reads \mbox{$\displaystyle{|\langle n^{(m)}|j^{(k)}\rangle|^2=1/2}$}. This means that one can realize a SIC POM in dimension 2 by a fuzzy measurement, with the corresponding diagonal Kraus operators 
\begin{align}\label{weak dim2}
A_1&={\rm diag} {\left({\textstyle\sqrt{\frac1{2}-\frac1{\sqrt{12}}},\sqrt{\frac1{2}+\frac1{\sqrt{12}}}}\right)},\nonumber\\  
A_2&={\rm diag} {\left({\textstyle \sqrt{\frac1{2}+\frac1{\sqrt{12}}},\sqrt{\frac1{2}-\frac1{\sqrt{12}}}}\right)},
\end{align}
followed by a measurement in a basis which is chosen in accordance with the result of the first measurement.
The solution indicates that the two bases of the second measurements must be unbiased to each other and to the computational basis! For example, if the $A$s of Eq.~\eqref{weak dim2} are diagonal in the $\sigma_3$ basis (where the $\sigma$s are the Pauli operators), then the two MUB of the second measurements could be the $\sigma_1$ basis and the $\sigma_2$ basis.  Actually all SIC POMs for a qubit are unitarily equivalent to the ``tetrahedron measurement'' (TM), whose outcomes correspond to four vectors that define a tetrahedron in the Bloch sphere \cite{berge04,renes04}. As we just showed, the TM could be realized in a two-step measurement process, for example, by using  a set-up similar to the one presented in Fig.~\ref{fig:sicpom_hw}.

Actually, the TM was successfully implemented in an optical system \cite{ling06}, where the qubit was encoded in a photon's polarization (``polarization qubit'') rather than in spatial alternatives (and therefore there was no need to stabilize interferometric loops in the set-up). The set-up of \cite{ling06} also consisted of a sequence of two measurements, quite analogous to what is described above. In that set-up, a partially polarizing beam splitter (PPBS) was used to implement the Kraus operators $A_k$ and then, depending on whether the photon was transmitted or reflected, a measurement of  $\sigma_1$ or $\sigma_2$ followed. 

Evidently, there is an operational relation between the TM and the three MUB in dimension 2. The latter are used to construct the former by means of successive measurements.  This relation actually stems from the common mathematical structure of the four kets (in the Hilbert space of a qubit) corresponding to the TM and the four kets composing the two bases unbiased to the computational basis and to each other. To see this more clearly, consider the columns of the following matrices:
\beq
\frac1{N}\left(%
\begin{array}{rr}
 \chi&\chi\\
 1&-1
\end{array}%
\right),\;\;
\frac1{N}\left(%
\begin{array}{rr}
  1&1\\
  i\chi&-i\chi
\end{array}%
\right).
\eeq
For  \mbox{$N=\sqrt{3+\sqrt{3}}$} and  \mbox{$\chi=\sqrt{2+\sqrt{3}}$}, the four columns represent the kets  corresponding to the TM. While for  \mbox{$N=\sqrt{2}$} and  \mbox{$\chi=1$} the two columns of each matrix form a basis. The two bases are unbiased to each other and to the computational basis. We will see below that similar  relations appear in dimensions 3, 4, and 8. 

Finally, since the TM is equivalent to a HW SIC POM, it could also be implemented by a two-step process: a measurement with the corresponding Kraus operators of Eq.~\eqref{a} followed by a measurement in the Fourier basis, Eq.~\eqref{b} (call it the $\sigma_1$ basis). We note that while the outcomes of the first measurement for this process and the outcomes for the fuzzy measurement in the  process discussed in this section are the same, the Kraus operators, and therefore the implementations, are different. 

\section{Dimension 3: A qutrit}\label{dim3}
According to the conditions listed in Eqs.~\eqref{cond1} and~\eqref{cond2}, the ansatz of Eqs.~\eqref{krauss}-\eqref{general pom} yields a SIC POM in dimension 3, if and only if  \mbox{$|\langle m|n^{(m)}\rangle|^2=1/3$} and, either \mbox{$\lambda=1/2$} and
\beq\label{lambda p}
\left|\!\bra{n^{(m)}}\!\big(1+\!\ket{m}\!\bra{m}\!+\!\ket{k}\!\bra{k}\big)\!\!\ket{j^{(k)}}\!\right|^2\!=1,
\eeq
with \mbox{$k\neq m$}, or \mbox{$\lambda=-1/2$} and
\beq\label{lambda m}
{\textstyle\left|\bra{n^{(m)}}l\right\rangle\left\langle{l}\ket{j^{(k)}}\right|}^2=\frac1{9},
\eeq
with \mbox{$k\neq m\neq l$}. Whereas Eq.~\eqref{lambda p} does not have a solution,  Eq.~\eqref{lambda m}  can be solved. One possible solution is \mbox{$|\langle n^{(m)}|l\rangle|^2=|\langle l|j^{(k)}\rangle|^2=1/3$}. This implies that a SIC POM in dimension 3 could be broken into a sequence of a fuzzy measurement with \mbox{$\lambda=-1/2$},
\beq\label{weak dim3}
A^\dagger_k=\frac1{\sqrt{2}}\bigl(\ket{k\oplus 1}\bra{k\oplus 1}+\ket{k\oplus 2}\bra{k\oplus 2}\bigr),
\eeq
with \mbox{$k=1,2,3$}, followed by a projective measurement onto a basis unbiased to the computational basis (in which the $A$s are diagonal).

Actually, in dimension 3, there exists a one-parameter family of non-equivalent HW SIC POMs \cite{renes04,appleby05,zauner11},
\begin{align}\label{sic3}
&\Pi_{k,j}=\frac1{9}\ket{\phi_{k,j}(\gamma)}\bra{\phi_{k,j}(\gamma)},\nonumber\\
&\ket{\phi_{k,j}(\gamma)}=\frac1{\sqrt{2}}\left(\ket{k}-e^{i2\gamma}\omega^j\ket{k\oplus 1}\right),
\end{align}
where \mbox{$k,j=1,2,3$}, \mbox{$0\leq\gamma\leq\pi/6$}, and \mbox{$\omega=\exp(2i\pi/3)$}. This continuum of SIC POMs could be realized using our ansatz in the following way: First, a fuzzy measurement, with the corresponding Kraus operators of Eq.~\eqref{weak dim3}, is carried out. Then, if the $k$th outcome is found the system goes through the diagonal unitary  transformation
\beq\label{unitary dim3}
U^\dagger_k=\ket{k}\bra{k}-\ket{k\oplus 1}e^{i2\gamma}\bra{k\oplus 1}+\ket{k\oplus 2}\bra{k\oplus 2}.
\eeq
And lastly, the system is subjected to a projective measurement onto a basis unbiased to the computational basis, say the Fourier basis; cf. Eq.~\eqref{b} with \mbox{$d=3$}.

\begin{figure}[t]
\centering
\includegraphics[scale=1.]{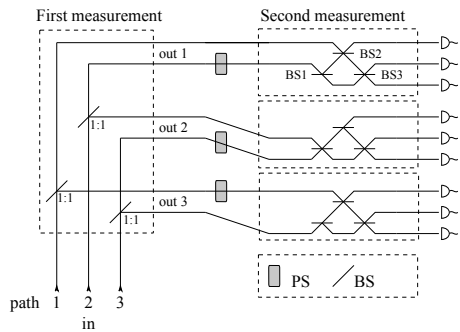}
\caption{An optical implementation of the one-parameter family of non-equivalent SIC POMs for a path qutrit.}
\label{fig:sicpom_dim3}
\end{figure}

This procedure implements the SIC POMs in Eq.~\eqref{sic3} for all $\gamma$. From an operational point of view, this result shows that the entire family of non-equivalent SIC POMs could, in principle, be realized in a single set-up. In Fig.~\ref{fig:sicpom_dim3} we present such an implementation in an optical setting for a path qutrit. The balanced (1:1) BSs in the first part are used to implement the fuzzy measurement, 	then the unitary transformations of Eq.~\eqref{unitary dim3} are implemented by path dependent PSs placed in the appropriate paths, and finally the Fourier transformation is applied to the state of the qutrit after which the path of the photon is detected. The Fourier transformation is implemented using three BSs, BS1, BS2, and BS3, which implement the transformations by the unitary operators \mbox{${\textstyle (\sigma_3+\sigma_1)/\sqrt{2}}$}, \mbox{${\textstyle (\sigma_3+\sqrt{2}\sigma_1)/\sqrt{3}}$}, and \mbox{${\textstyle (\sigma_3+\sigma_2)/\sqrt{2}}$}, respectively.

In the above construction we chose the Fourier basis for the second measurement independent of the outcome of the first measurement. However, the same family of SIC POMs (or its unitarily equivalent) could be realized with a different choice for the basis for the second measurement, as long as Eq.~\eqref{lambda m} is obeyed. For example, one may use the three MUB which are also unbiased to the computational basis as the bases for the second measurement.

\section{Dimension 4: Two qubits}\label{dim4}
In dimension $4$, there is only one known HW SIC POM, and all the other known SIC POMs are equivalent to it \cite{appleby05}. This SIC POM is composed of 16 subnormalized projectors onto 16 (fiducial) kets. The latter are represented in the following matrices  as columns  with \mbox{$N=\sqrt{5+\sqrt{5}}$} and {$\chi=\sqrt{2+\sqrt{5}}$} \cite{bengtsson10},%
\begin{alignat}{2}\label{fiddim4}
&\frac1{N}{\left(%
\begin{array}{rrrr}
\chi&\chi&\chi&\chi\\
1&-1&1&-1\\
1&1&-1&-1\\
1&-1&-1&1
\end{array}%
\right)}\!,
&\;&\frac1{N}{\left(%
\begin{array}{rrrr}
1&1&1&1 \\
1&-1&1&-1\\
i\chi&i\chi&-i\chi&-i\chi\\
-i&i&i&-i
\end{array}%
\right)}\!,\nonumber\\
&\frac1{N}{\left(%
\begin{array}{rrrr}
1&1&1&1 \\
i\chi&-i\chi&i\chi&-i\chi\\
i&i&-i&-i\\
-1&1&1&-1
\end{array}%
\right)}\!,
&\;&\frac1{N}{\left(%
\begin{array}{rrrr}
1&1&1&1 \\
i&-i&i&-i\\
1&1&-1&-1\\
-i\chi&i\chi&i\chi&-i\chi
\end{array}%
\right)}\!.
\end{alignat}%
Each of these matrices could be written as a diagonal matrix times a unitary matrix. The set of bases, corresponding to each unitary matrix, together with the computational basis, form the complete set of MUB in dimension 4. To be more specific, the diagonal matrices are
\begin{align}\label{Adim4}
A_1=&\frac1{N}{\rm diag}(\chi,1,1,1),\;\;A_3=\frac1{N}{\rm diag}(1,1,\chi,1),\nonumber\\
A_2=&\frac1{N}{\rm diag}(1,\chi,1,1),\;\;A_4=\frac1{N}{\rm diag}(1,1,1,\chi),
\end{align}
and the  unitary matrices are
\begin{alignat}{2}\label{Bdim4}
{\cal U}_1\!&=\!\frac1{2}{\left(%
\begin{array}{rrrr}
1&1&1&1 \\
1&-1&1&-1\\
1&1&-1&-1\\
1&-1&-1&1
\end{array}%
\right)}\!,\;
&{\cal U}_3\!&=\!\frac1{2}{\left(%
\begin{array}{rrrr}
1&1&1&1 \\
1&-1&1&-1\\
i&i&-i&-i\\
-i&i&i&-i
\end{array}%
\right)}\!,\nonumber\\
{\cal U}_2\!&=\!\frac1{2}{\left(%
\begin{array}{rrrr}
1&1&1&1 \\
i&-i&i&-i\\
i&i&-i&-i\\
-1&1&1&-1
\end{array}%
\right)}\!,\;
&{\cal U}_4\!&=\!\frac1{2}{\left(%
\begin{array}{rrrr}
1&1&1&1 \\
i&-i&i&-i\\
1&1&-1&-1\\
-i&i&i&-i
\end{array}%
\right)}\!.
\end{alignat}%
Noting that \mbox{$\sum_jA^\dagger_jA_j=1$}, we identify the $A$s with the Kraus operators of a measurement.

Actually, the operations of Eq.~\eqref{Bdim4} transform the computational basis into the MUB, 
\begin{align}\label{MUBdim4}
{\mathfrak B}_1&=\left\{\begin{array}{r}
\frac1{\sqrt{2}}(\ket{0}+\ket{1})\\
\frac1{\sqrt{2}}(\ket{0}-\ket{1})\\
\end{array}\right\}\otimes\left\{\begin{array}{r}
\frac1{\sqrt{2}}(\ket{0}+\ket{1})\\
\frac1{\sqrt{2}}(\ket{0}-\ket{1})\\
\end{array}\right\},\nonumber\\
{\mathfrak B}_2&=\left\{\begin{array}{r}
\frac1{\sqrt{2}}(\ket{0}+i\ket{1})\\
\frac1{\sqrt{2}}(\ket{0}-i\ket{1})\\
\end{array}\right\}\otimes\left\{\begin{array}{r}
\frac1{\sqrt{2}}(\ket{0}+i\ket{1})\\
\frac1{\sqrt{2}}(\ket{0}-i\ket{1})\\
\end{array}\right\},\nonumber\\
{\mathfrak B}_3&={\rm CZ}\left\{\begin{array}{r}
\frac1{\sqrt{2}}(\ket{0}+i\ket{1})\\
\frac1{\sqrt{2}}(\ket{0}-i\ket{1})\\
\end{array}\right\}\otimes\left\{\begin{array}{r}
\frac1{\sqrt{2}}(\ket{0}+\ket{1})\\
\frac1{\sqrt{2}}(\ket{0}-\ket{1})\\
\end{array}\right\},\nonumber\\
{\mathfrak B}_4&={\rm CZ}\left\{\begin{array}{r}
\frac1{\sqrt{2}}(\ket{0}+\ket{1})\\
\frac1{\sqrt{2}}(\ket{0}-\ket{1})\\
\end{array}\right\}\otimes\left\{\begin{array}{r}
\frac1{\sqrt{2}}(\ket{0}+i\ket{1})\\
\frac1{\sqrt{2}}(\ket{0}-i\ket{1})\\
\end{array}\right\},
\end{align}
where CZ stands for the controlled-Z (phase flip) operation, \mbox{${\rm CZ}={\rm diag}(1,1,1,-1)$}. The  bases ${\mathfrak B}_1$ and ${\mathfrak B}_2$ are composed of product states, while the bases ${\mathfrak B}_3$ and ${\mathfrak B}_4$ consist of maximally entangled states.

The structural relation between the fiducial kets, Eq.~\eqref{fiddim4}, and  the kets that compose the four MUB in dimension 4, Eq.~\eqref{Bdim4}, is now clear. For \mbox{$N=\sqrt{5+\sqrt{5}}$} and \mbox{$\chi=\sqrt{2+\sqrt{5}}$}, the columns of each matrix in  Eq.~\eqref{fiddim4} form  the 16 fiducial kets. While for \mbox{$N=1/2$} and \mbox{$\chi=1$}, the columns of each matrix form a basis. These bases are mutually unbiased to each other and to the computational basis, cf. Eq.~\eqref{Bdim4}.

The structure of the fiducial kets  of Eq.~\eqref{fiddim4} (which form the SIC POM in dimension 4) allows us to implement the SIC POM by two successive measurements: a measurement whose Kraus operators are listed in Eq.~\eqref{Adim4} and, depending on the measurement outcome, a measurement in one of the MUB of Eq.~\eqref{MUBdim4}. We should not fail to mention that the Kraus operators of Eq.~\eqref{Adim4} correspond to a fuzzy measurement with $\lambda=1/\sqrt{5}$, cf. Eq.~\eqref{fuzzy}.

A proposal for implementing this SIC POM for a 2-qubit system, by successive measurements was recently given in \cite{letter}. This proposal is based on the ideas presented in \cite{englert01} where 2-qubit states are encoded in a single-photon state. In this scheme the photon's polarization is one qubit, the polarization qubit, and the orbital alternative of the photon is the second qubit, the path qubit. Performing a unitary transformation on the 2-qubit state, amounts to sending the photon through a set of passive linear optical elements (optical plates) that unitarily change the state of the path and polarization qubits \cite{englert01}. In particular, in order to realize the fuzzy measurement, two more path-qubits were used as ancillae. In each path, the parameters of the different optical elements were set such that a photon which enters the apparatus with a polarization-path statistical operator $\rho$, exits at at path $k$  with the 2-qubit statistical operator \mbox{$A_k\rho A_k^\dagger/\tr{A_k\rho A_k^\dagger}$}. Thus, a projective measurement (with 4 possible outcomes) on the ancillary qubits effectively produces the desired POM on the 2-qubit system $\{A_k\}$. At the second step, measurements in the MUB were implemented as described in \cite{letter}.

\section{Dimension 8: Three qubits}\label{dim8}
In dimension 8 there are three known non-equivalent SIC POMs. One of them is covariant with respect to the three-qubit Pauli group \cite{zauner11,hoggar}, while the other two are covariant with respect to the HW group \cite{scott10}. According to the result presented in Sec.~\ref{general}, the latter two could be realized by a diagonal-operator measurement followed by a measurement in the Fourier basis of the three qubits. Interestingly, as we see in what follows, the former SIC POM (also known as `Hogger's SIC POM' \cite{hoggar}) could be broken into a diagonal-operator measurement followed by projective measurements in eight MUB, similarly to what happens in dimensions 2, 3, and 4.  

Hogger's SIC POM is composed of (subnormailzed) projectors onto 64 kets. The latter are constructed from the action of the three-qubit Pauli group elements on a fiducial ket $\ket{\phi}$,
\beq\label{64}
\ket{ ^{(k,l,m)}_{(n,r,s)}\!}=Z_1^{n}X_1^{k}\otimes Z_2^{r}X_2^{l}\otimes Z_3^{s}X_3^{m}\ket{\phi},
\eeq
where all indices take on the values 1, 2. Here \mbox{$Z=\sigma_3$} and \mbox{$X=\sigma_1$} are the generators of the Pauli group  in dimension 2, and their subscripts in Eq.~\eqref{64} label the degree of freedom on which they act. In what follows we omit this subscript when no ambiguity arises. We refer to the basis in which $\sigma_3$ is diagonal as the computational basis. In this basis, the fiducial ket $\ket{\phi}$ is represented by  
\beq
\phi=\frac1{\sqrt{6}}(\sqrt{2},0,-\omega^*,\omega^*,-\omega,-\omega^*,0,0)^\intercal,
\eeq
where $\omega=\sqrt{i}$ is the fundamental eighth root of unity. 

This SIC POM can be broken into two successive measurements: The Kraus operators corresponding to the first 
measurement are
\begin{align}
A_1&=\frac{2}{\sqrt{3}}{\rm diag}(0,-i\sqrt{2},\omega^*,i\omega^*,-\omega^*,-i\omega,0,0),\nonumber\\
A_2&=\frac{2}{\sqrt{3}}{\rm diag}(-\omega^*,\omega^*,-i\sqrt{2},0,0,0,i\omega,i\omega^*),\nonumber\\
A_3&=\frac{2}{\sqrt{3}}{\rm diag}(\omega^*,i\omega^*,0,\sqrt{2},0,0,-i\omega^*,\omega),\nonumber\\
A_4&=\frac{2}{\sqrt{3}}{\rm diag}(-\omega,-\omega^*,0,0,-i\sqrt{2},0,-i\omega^*,-i\omega^*),\nonumber\\
A_5&=\frac{2}{\sqrt{3}}{\rm diag}(-\omega^*,i\omega,0,0,0,\sqrt{2},-i\omega^*,\omega^*),\nonumber\\
A_6&=\frac{2}{\sqrt{3}}{\rm diag}(0,0,i\omega,i\omega^*,i\omega^*,i\omega^*,\sqrt{2},0),\nonumber\\
A_7&=\frac{2}{\sqrt{3}}{\rm diag}(0,0,i\omega^*,\omega,-i\omega^*,\omega^*,0,i\sqrt{2}),\nonumber\\
A_8&=\frac{2}{\sqrt{3}}{\rm diag}(\sqrt{2},0,-\omega^*,\omega^*,-\omega,-\omega^*,0,0).
\end{align}
The basis for the second measurement is chosen in accordance with the result of the first measurement. The eight different bases for the second measurement, together with the computational basis, form a complete set of MUB in dimension 8. This set is constructed as follows. Consider the two unbiased bases in dimension 2 --- the ones that correspond to the Pauli matrices $\sigma_1$ and $\sigma_2$. We label these bases by $\{\ket{e_v^b}\}$ where $v=1,2$ labels the vector in basis $b=1,2$. Furthermore, consider the operator
\begin{align}\label{mub gen}
{\cal G}(k,l,m)=\frac{1}{2}\Big(&1+\sigma_3^{k}\otimes\sigma_3^{l}\otimes\sigma_3^{m}+\sigma_3^{1-k}\otimes\sigma_3^{1-l}\otimes\sigma_3^{1-m}\nonumber\\
&-\sigma_3\otimes\sigma_3\otimes\sigma_3\Big),
\end{align}
with \mbox{$k,l,m=1,2$}. The states defined by a fixed triplet $(k,l,m)$,
\beq
\ket{e^{(k,l,m)}_{(n,r,s)}}={\cal G}(k,l,m)\ket{e_{n}^{m}}\otimes\ket{e_{r}^{k}}\otimes\ket{e_{s}^{l}},
\eeq
form a basis, while bases with different $(k,l,m)$ triplets are mutually unbiased. These bases together with the computational basis form a complete set of MUB for three qubits. The 64 fiducial kets in Eq.~\eqref{64} are
\beq
\ket{^{(k,l,m)}_{(n,r,s)}\!}=A_{(k,l,m)}\ket{e^{(k,l,m)}_{(n,r,s)}},
\eeq
where the index of the Kraus operator is written in binary representation, where $m$ is the least significant bit. Indeed, the last equation implies that Hogger's SIC POM can be realized by a measurement with the corresponding Kraus operators $A_{(k,l,m)}$ and, depending on the result, followed by a measurement in one of the MUB.

Finally, we note that the above construction of the SIC POM in dimension 8 is different in two points from the constructions given for the SIC POMs in dimensions 2, 3, and 4. First, the SIC POMs in dimensions 2, 3, and 4 are covariant with respect to the HW group while the Hogger's SIC POM is covariant with respect to the three-qubit Pauli group. And second, in dimensions 2, 3, and 4 the first measurement is a fuzzy measurement while in dimension 8 this is not  the case.

\section{Conclusions}\label{conclusions}
SIC POMs are considered to be hard to implement. Here, we are proposing to implement them by breaking the measurement process into two steps, having in mind that each step should be rather easy to implement. Based on this idea, we presented a systematic procedure that implements HW SIC POMs in finite-dimensional systems. The implementation is accomplished by a diagonal-operator measurement with high-rank outcomes followed by a rank-1 measurement in the Fourier basis.  As an example we have considered the realization of HW SIC POMs for a path qudit encoded in a single photon. Moreover, we found that if we take the first measurement to be a fuzzy measurement and we let the bases for the second measurement to be chosen in accordance with the result of the first measurement, then in the particular studied cases (dimensions 2, 3, and 4) an operational link between SIC POMs and MUB appears, that is, the MUB are used to implement the SIC POMs in the successive measurement scheme. A similar link was found in dimension 8 as well, but here the first measurement was not of the fuzzy kind.

There is still an open question as to the generality of such a relation and its origin. Currently it is unclear whether the successive measurement approach will provide a reasonable scheme for implementing SIC POMs in arbitrary dimensions and thus reveal their structure in high-dimensional Hilbert spaces.
\begin{acknowledgments}
We would like to thank Huangjun Zhu for valuable and stimulating discussions.
Centre for Quantum Technologies is a Research Centre of Excellence funded by Ministry of Education and National Research Foundation of Singapore. 
\end{acknowledgments}

\end{document}